\documentclass[3p]{elsarticle}
\pdfoutput=1
\usepackage[utf8]{inputenc}
\usepackage{hyperref,amsmath,amsfonts,natbib,graphicx,bm,subfig,floatrow,multirow,wrapfig,xfrac}
\usepackage[dvipsnames]{xcolor}
\usepackage{tabularx}
\usepackage{multicol}
\usepackage{comment}
\usepackage[linesnumbered,ruled,vlined]{algorithm2e}
\usepackage{psfrag}
\usepackage{placeins}

\newcommand{\rmd}{\ensuremath{\mathrm{d}}}
\newcommand{\der}[2]{\ensuremath{\dfrac{\rmd #1}{\rmd #2}}}
\newcommand{\pder}[2]{\ensuremath{\dfrac{\partial #1}{\partial #2}}}

\newcommand{\ang}[1]{\ensuremath{\left\langle #1 \right\rangle}}

\DeclareMathSymbol{\shortminus}{\mathbin}{AMSa}{"39}

\graphicspath{{figures/}}

\journal{Physical Review Fluids}

\begin{document}
\begin{frontmatter}

\title{Equation-informed data-driven identification of flow budgets and dynamics}

\author[SU]{Nataliya Sevryugina\corref{corresponding}}
\cortext[corresponding]{Corresponding author}
\ead{nataliya.sevryugina@students.unibe.ch}
\author[SU]{Serena Costanzo}
\author[MArm]{Stephen de Bruyn Kops}
\author[CAM]{Colm-cille Caulfield}
\author[CNAM]{Iraj Mortazavi}
\author[SU,CNAM]{Taraneh Sayadi}

\address[SU]{Sorbonne Université, Institut Jean Le Rond $\partial$'Alembert, UMR 7190, 4 Place Jussieu, 75252 Paris Cedex 05, France}
\address[CAM]{Department of Applied Mathematics and Theoretical Physics, University of Cambridge, Cambridge CB3 0WA, UK}
\address[MArm]{Department of Mechanical and Industrial Engineering, University of Massachusetts Amherst, Amherst, MA, USA, 01003}
\address[CNAM]{Mathematical and Numerical Modelling Laboratory, Conservatoire National Arts et Métiers, F-75003 Paris, France}

\begin{abstract}
Computational Fluid Dynamics (CFD) is an indispensable method of fluid modelling in engineering applications, reducing the need for physical prototypes and testing for tasks such as design optimisation and performance analysis. Depending on the complexity of the system under consideration, models ranging from low to high fidelity can be used for prediction, allowing significant speed-up. However, the choice of model requires information about the actual dynamics of the flow regime. Correctly identifying the regions/clusters of flow that share the same dynamics has been a challenging research topic to date. In this study, we propose a novel hybrid approach to flow clustering. It consists of characterising each sample point of the system with equation-based features, i.e. features are budgets that represent the contribution of each term from the original governing equation to the local dynamics at each sample point. This was achieved by applying the Sparse Identification of Nonlinear Dynamical systems (SINDy) method pointwise to time evolution data. The method proceeds with equation-based clustering using the Girvan-Newman algorithm. This allows the detection of communities that share the same physical dynamics. The algorithm is implemented in both Eulerian and Lagrangian frameworks. In the Lagrangian, i.e. dynamic approach, the clustering is performed on the trajectory of each point, allowing the change of clusters to be represented also in time. The performance of the algorithm is first tested on a flow around a cylinder. The construction of the dynamic clusters in this test case clearly shows the evolution of the wake from the steady state solution through the transient to the oscillatory solution. Dynamic clustering was then successfully tested on turbulent flow data. Two distinct and well-defined clusters were identified and their temporal evolution was reconstructed. 
\end{abstract}

\end{frontmatter}
\section{Introduction}

With advances in computing power, larger and more accurate simulations are being performed, capable of capturing detailed interactions of various physical phenomena present in the flow. However, the cost of these simulations remains prohibitive, especially when parametric dependencies need to be explored. One solution to reduce the cost of such simulations is to rely on different modelling strategies of variable fidelity depending on the underlying dynamics present. Such decompositions have been attempted in the past, for example by combining low and high fidelity simulations, i.e. hybrid RANS-LES simulations \cite{rans_les,rans_les_2}, wall models to model near wall behaviour eliminating the restrictive resolutions otherwise needed \cite{WMLES}, and variable fidelity reduced order models \cite{DGDD}. However, the main challenge is how to make this regional clustering automatic rather than user-specified. 

Given recent advances in machine learning and the availability of data, researchers have naturally turned to data-driven techniques for clustering flow regimes. This data-driven approach allows the automatic identification of dynamically distinct regions for the application of appropriate modelling techniques. However, purely data-driven techniques for dynamic identification are sometimes considered to be a black-box approach (fidelity was described by Wiener \cite{Wiener1948}), as the dynamics are identified taking into account the response of the overall flow process. Thus, the interpretability of the results is overlooked. In order to increase the interpretability and accuracy, it seems much more constructive to find a way to incorporate some level of physical knowledge into the dynamics identification process. To this end, Brunton \textit{et. al.}~\cite{Brunton2016} suggested that this wealth of data could be used to derive the equations governing the existing dynamics, which can be considered as an alternative gray box approach, as some information about the underlying physics is included in the identification process. Other researchers have also explored similar identification processes using data \cite{ronaalp_2}, \cite{ronaalp_1}, \cite{ML_for_cfd}, \cite{ML_for_cfd_2}.

These gray-box models can also be integrated into control applications, as they identify the system by analysing the relationship between its inputs and outputs. Linear system identification \cite{Webster2017} has been used in control in the past \cite{Herve2012,Huang2008}, but linear approaches often lead to instability when applied to transient measurements of nonlinear problems \cite{Huang2008} commonly encountered in flow applications. For such transient and nonlinear environments, nonlinear algorithms are more appropriate. One such nonlinear algorithm is SINDy~\cite{Brunton2016}, which is capable of identifying nonlinear models from data. In SINDy, complex nonlinear dynamics are represented as a linear combination of simpler nonlinear functions. These functions (feature space) typically include terms inspired by the underlying governing equations, allowing the algorithm to incorporate relevant physical information \cite{loiseau2018constrained}. This method has also been applied to nonlinear flow problems \cite{loiseau2018sparse} and shows great promise. \\

However, what we propose in this paper differs from pure equation identification as proposed in SINDy and its variants. Here, the aim is to cluster the important dynamics of the system by identifying the dominant terms of the governing equation and then clustering the regions accordingly. In other words, we do not aim to learn the governing equations, but rather the terms that are most active in the overall dynamics. Our approach follows what was originally proposed by \cite{callaham2021learning} and is based on the idea that many regimes described by complex partial differential equations are governed by only a few nonlinear terms, allowing the creation of simpler models for specific dynamics of the system. Identifying and locating the different dynamics present in complex physical systems, and ultimately understanding how they affect the overall solution, is essential for the next step, which is to select the most appropriate model to predict the overall flow. In other areas, such as reactive flows, this idea has been successfully used to cluster different flow regimes based on the level of detailed chemical mechanisms required to represent the interaction of flow and chemistry~\cite{reactive_flow}. In this study, unlike \cite{callaham2021learning}, we use graph clustering, which has been shown to be suitable for such problems in the past \cite{ronaalp_1}, \cite{ronaalp_2}, \cite{costanzo2022}. More importantly, we propose a novel algorithm to dynamically evolve the clusters, moving from a conventional Eulerian framework used in such identification algorithms to a Lagrangian alternative. This would make the algorithm more suitable for intermittent problems or unsteady transient problems where the significant dynamic regions might evolve over time. \\

The paper is structured as follows. Section~\ref{sec:BIA} describes the basics of the identification algorithm in a Eulerian framework, where the sample points are kept stationary in physical space and the temporal dynamics are discovered locally at each point, leading to the algorithm termed BIA. The performance of the algorithm is evaluated using an academic test case of flow around a cylinder. Section~\ref{sec:dyn-bia} then discusses the modification of the algorithm by considering a Lagrangian frame of reference, where each sample point is allowed to move in the physical domain and the inference is performed in this moving frame of reference for each point, leading to a dynamic version of the algorithm; i.e. Dynamic-BIA. Following this strategy, the dynamically distinct regions of the flow are allowed to evolve as the flow evolves. Finally, in section~\ref{sec:turbulence}, the dynamic algorithm is applied to a turbulent stratified flow to identify different regions of dynamic significance. Conclusions and future work are then discussed in section~\ref{sec:conclusions}. 
\section{Budget identification algorithm (BIA) -- Eulerian framework}
\label{sec:BIA}

This section contains the summary of the mathematical background of the BIA algorithm, first presented in \cite{costanzo2022}. The algorithm is divided in two steps: (i) regression and (ii) clustering, which are presented in this section. 

\subsection{Regression}
\label{sec:Regression}
Similar to the algorithm proposed by Brunton \textit{et al.}~\cite{Brunton2016}, the equation-based identification model leverages advances in machine learning and sparsity techniques to discover the governing equations from data measurements, without having any a priori information on the form of the expected model. The main difference to the original approach is that in our regression algorithm the terms included in the library of features are only the terms of the governing equation and are already known, and no extra terms are supplied. In what follows the sparse regression algorithm is first summarized and then applied to a data of a flow around a cylinder. 

The generic nonlinear dynamical system can be written as:
\begin{equation}
\der{\boldsymbol{y}(t)}{t}=\boldsymbol{f}(\boldsymbol{y}(t)) \text, 
\label{eq: PDE1}
\end{equation}
where $\boldsymbol{y}(t) \in \mathbb{R}^{n}$ is time dependent data and $\boldsymbol{f}(\boldsymbol{y}(t))$ denotes general nonlinear function. Since $\boldsymbol{y}(t)$ is assumed known, the aim is therefore to determine the form of the unknown generic nonlinear term $\boldsymbol{f}(\boldsymbol{y}(t))$ that would satisfy Eq.~\ref{eq: PDE1}. We proceed by expressing the terms of Eq.~\ref{eq: PDE1} in matrix form. The time evolution of the data, collected at different time instants $i=1,\dots,n_{t}$ is stored in $\mathbf{Y}\in \mathbb{R}^{n_{t}\times n}$. The second matrix $\dot{\mathbf{Y}}$ contains the time derivatives of $\mathbf{Y}$ at the same time instants. The derivative information is usually not available and can be computed using any of the numerical schemes such as the finite difference scheme.
\begin{equation}
\mathbf{Y} = \begin{bmatrix}
\boldsymbol{y}^{T}(t_{1})\\
\boldsymbol{y}^{T}(t_{2})\\
\vdots\\
\boldsymbol{y}^{T}(t_{n_{t}})\\
\end{bmatrix}\quad \text, 
\quad\quad\quad
\dot{\mathbf{Y}} = \begin{bmatrix}
\dot{\boldsymbol{y}}^{T}(t_{1})\\
\dot{\boldsymbol{y}}^{T}(t_{2})\\
\vdots\\
\dot{\boldsymbol{y}}^{T}(t_{n_{t}})\\
\end{bmatrix}\text .
\label{eq: matrix data sindy}
\end{equation}

The right-hand side of Eq.~\ref{eq: PDE1} can be expressed as multiplication of $\mathbf{Y}$ by a set of linear and nonlinear functions, also called a candidate library $\boldsymbol{\Theta}(\mathbf{Y})$. The functions can assume any mathematical form, thus $\boldsymbol{\Theta}(\mathbf{Y})$ can be built from polynomials of any order or trigonometric terms. For PDEs, it can also include partial derivatives and external forcing. An example of $\boldsymbol{\Theta}(\mathbf{Y})$ could be: 

\begin{equation}
\boldsymbol{\Theta}(\mathbf{Y}) = \begin{bmatrix}
	| & | & | &  & |&\\
	1& \mathbf{Y}& \mathbf{Y}\mathbf{Y}_y&\cdots& \mathbf{Y}_y\mathbf{Y}_x& \cdots\\
	| & | & | &  & |&\\
	\end{bmatrix} \text .
\label{eq: candidate library}
\end{equation}

\noindent Finally, the nonlinear governing Equation~\eqref{eq: PDE1} is rewritten as:
\begin{equation}
\dot{\mathbf{Y}} = \boldsymbol{\Theta}(\mathbf{Y}) \boldsymbol{\Xi}, 
\label{eq: PDE_candidate}
\end{equation}
where $\boldsymbol{f}(\boldsymbol{y}(t))$ is defined as a linear combination of the nonlinear candidate functions from $\boldsymbol{\Theta}(\mathbf{Y})$ and $\boldsymbol{\Xi}=\left[\xi_{1}\, \xi_{2}\,\cdots \, \xi_{n}\right]$. The latter represents the coefficients $\xi_{i}$ of the linear combination stored in matrix form. 

It is reasonable to assume that, at any particular point in space and time, specific dynamics of the system are affected only by a few of these nonlinear terms. Those terms represent the nonzero coefficients in $\boldsymbol{\Xi}$ and can be identified by any sparse regression algorithm. One of them might be the LASSO algorithm~\cite{tibshirani1996regression}:

\begin{equation}
\boldsymbol{\xi}=\text{arg}\underset{\mathbf{\xi}'}{\text{min}} \| \boldsymbol{\Theta}\boldsymbol{\xi}' - \dot{\boldsymbol{y}}\|_{2} + \lambda \| \boldsymbol{\xi}' \|_{1},
\label{eq: LASSO}
\end{equation}
where the solution is penalized in $\ell_{1}$ to promote sparsity. Identifying active terms and at the same time enhancing the sparsity of the solution can be done by applying other algorithms such as sequentially thresholded least squares (STLSQ)~\cite{zhang2019convergence}, the sparse relaxed regularized regression (SR3)~\cite{zheng2018unified}, the stepwise sparse regression (SSR)~\cite{boninsegna2018sparse} and Bayesian approaches~\cite{pan2015sparse}.\\

\noindent When the matrix $\boldsymbol{\Xi}$ is calculated, the model for the evolution equations~\eqref{eq: PDE1} can be expressed as:
\begin{equation}
\dot{\boldsymbol{y}} = \boldsymbol{\Xi}^{T}\left( \boldsymbol{\Theta}(\boldsymbol{y}^{T})\right)^{T} \text .
\label{eq: Sindy model}
\end{equation}

The steps are also presented in Algorithm~\ref{algo: sindy}. This algorithm has been used in different applications such as PDE identification ~\cite{rudy2017data,rudy2019data,kaheman2020sindy}, model reduction~\cite{loiseau2018sparse,loiseau2018constrained,loiseau2021pod}, dynamics identification~\cite{callaham2021learning}, and control~\cite{kaiser2018sparse}. 

\begin{algorithm}[H]
\label{algo: sindy}
\caption{Budget Identification Algorithm (BIA) - regression}
\KwIn{Data matrix $Y \in \mathbb{R}^{n_t \times n \times points}$ with $n_t$ timesteps and $n$ features, time derivative matrix $\dot{Y} \in \mathbb{R}^{n_t \times n \times points}$, library of candidate functions $\Theta(Y)$, threshold $\lambda$, regularization parameter $\alpha$}
\KwOut{Sparse coefficient matrix $\Xi$}

\Begin{
 1. Select set of $n_{eq}$ functions that represent terms of physical equation that would represent function library\;
 \For {each $i$ in $points$}{
    2. Define $Y_i\in \mathbb{R}^{n_t \times n}$ and construct $\dot{Y}_i\in \mathbb{R}^{n_t \times n}$\ if not already available with numerical schemes\;
    3. Apply previously created candidate function library on  $Y_i$ thus constructing  candidate function library matrix $\Theta(Y_i)  \in \mathbb{R}^{n_t \times n_{eq}}$ \;
    4. Formulate the nonlinear system as $\dot{Y_i} = \Theta(Y_i) \Xi_i$\;

    5. \textbf{Initial Least Squares:} Solve for $\Xi_i$ using ridge regression: 
    \[
    \Xi_i \gets \arg\min_{\Xi_i} \|\dot{Y}_i - \Theta(Y_i) \Xi_i \|^2_2 + \alpha \|\Xi_i\|^2_2
    \]
    
    \While {coefficients in $\Xi_i$ are larger than threshold $\lambda$}{
        6. Introduce sparsity by setting coefficients below $\lambda$ in $\Xi_i$ to zero\;
        7. Recompute $\Xi_i$ using Ridge Regression on non-zero terms:
        \[
        \Xi_i \gets \arg\min_{\Xi_i} \|\dot{Y_i} - \Theta(Y_i) \Xi_i \|^2_2 + \alpha \|\Xi_i\|^2_2
        \]
    }
    Accumulate each solved $\Xi_i \in \mathbb{R}^{n_{eq}}$ in $\Xi$
    }

    7. \Return $Xi\in \mathbb{R}^{points \times n_{eq}}$ that is the input necessary for clustering with the Algorithm~\ref{algo: newman};
}

\end{algorithm}

\subsubsection{Regression applied to flow around a cylinder}

\begin{figure}[tb]
\centering
\includegraphics[width=0.7\linewidth]{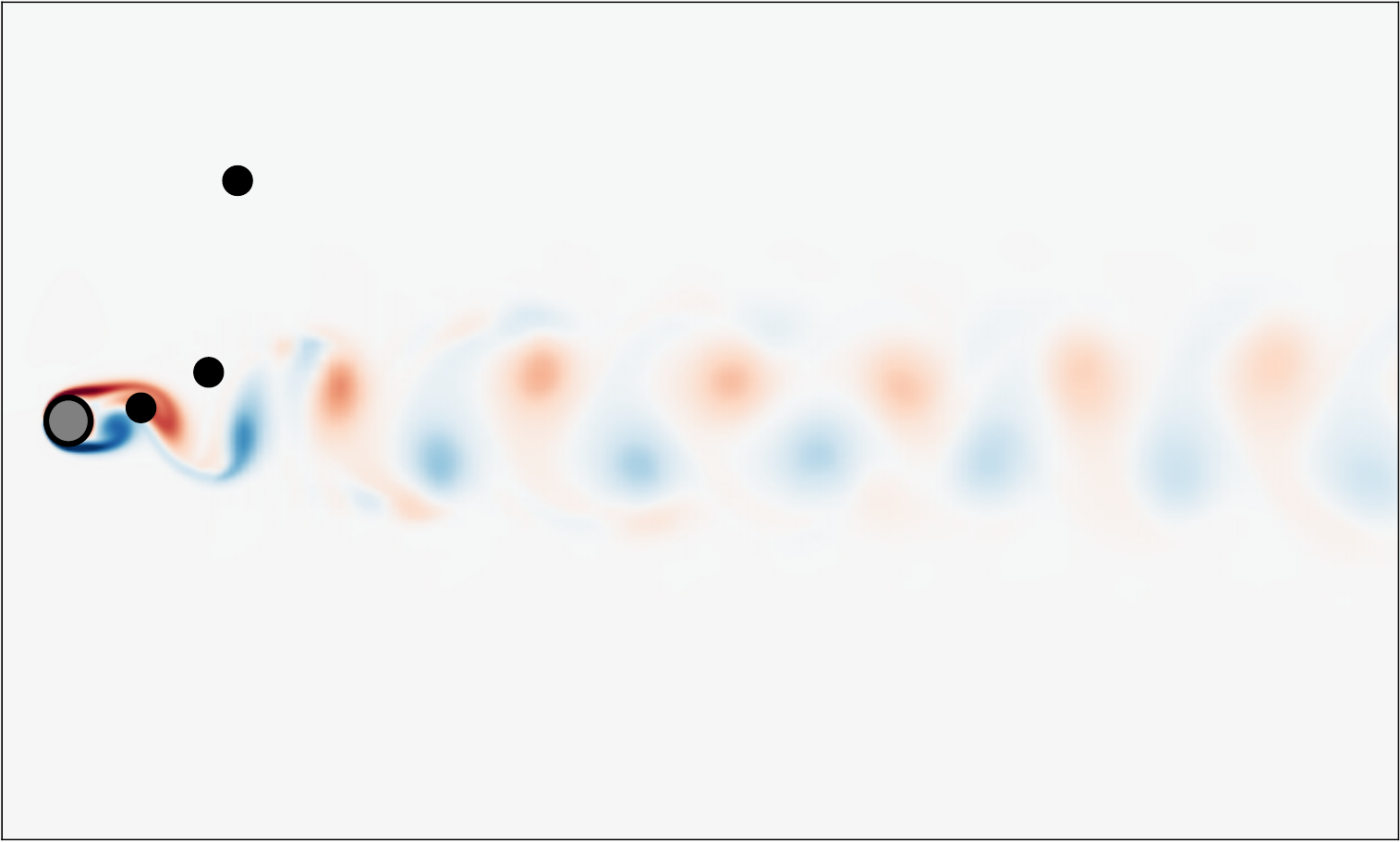}%
\caption{Vorticity field for a flow around a cylinder, $Re=200$, selection of three random points in the domain that would represent different flow regimes \cite{costanzo2022}.}
\label{fig: vorticity and points}
\end{figure}

The algorithm is assessed by applying it to the two-dimensional flow around a cylinder at $Re=200$. In the Figure~\ref{fig: vorticity and points} the vorticity field defined within the domain $-2\leq x \leq 30$ and $-9\leq y \leq 9$ is represented. The three black points at different distances from the wake generated behind the cylinder are selected to test the regression step of the BIA algorithm and compare the active terms of each point.

The discrete data is collected over $n_{t}=340$ time steps, which covers one full shedding cycle. The grid is refined around the cylinder and its wake, with the cylinder’s surface defined by Lagrangian points. In this grid, pressure is located at the center of each cell, while velocity is placed along the edges. To align all variables at the same location, the velocity is interpolated to the center of each cell before selecting points. A subset $\mathcal{I}$ of $N_{p}=3000$ points is selected from the domain, excluding points inside the cylinder and on its surface. The points are spaced twice the size of the smallest grid cells, with more points placed in areas where the grid is finer.

For each of the points, the temporal-spatial derivatives are numerically approximated and the library of candidate functions for each point is defined as:
\begin{equation}
\begin{aligned}
\boldsymbol{\Theta}(\dot{\mathbf{y}}^{(j)}) = &\left[ \quad p_{x}^{(j)}, p_{y}^{(j)},  \right.\\
&\quad\,\,\, \dfrac{1}{Re}u_{xx}^{(j)}, \dfrac{1}{Re}u_{yy}^{(j)}, \dfrac{1}{Re}v_{xx}^{(j)},\dfrac{1}{Re}v_{yy}^{(j)},  \\
&\quad\, \left. uu_{x}^{(j)},  vu_{y}^{(j)},  uv_{x}^{(j)}+ vv_{y}^{(j)}\quad \right] \text .
\end{aligned}
\label{eq : library candidates}
\end{equation}
\noindent where the subscript indicates the corresponding partial derivative, for $j=1,\dots, N_{p}$ points. The temporal evolution of the flow at each point can be rewritten as:
\begin{equation}
\mathbf{y}^{(j)}= \boldsymbol{\Theta}(\dot{\mathbf{y}}^{(j)}) \boldsymbol{\Xi}^{(j)} \text , 
\end{equation}
where $\boldsymbol{\Xi}^{(j)}$ are the coefficients of the polynomial combination for the generic point $j\in \mathcal{I}$.
The active terms of the polynomial combination are identified by applying a STRidge introduced in~\cite{rudy2017data} and adapted for group sparsity. 
\begin{equation}
\boldsymbol{\xi}=\text{arg}\underset{\mathbf{\xi}'}{\text{min}} \| \boldsymbol{\Theta}\boldsymbol{\xi}' - \dot{\boldsymbol{y}}\|_{2} + \lambda \| \boldsymbol{\xi}' \|_{2} \text , 
\label{eq: RidgeRegression}
\end{equation}
In this algorithm, the classic Ridge regression shown in Eq.~\ref{eq: RidgeRegression} is followed by a recursive penalization in $\ell_{0}$, removing coefficients smaller than a certain threshold to avoid overfitting.

\begin{figure}[tbh]
\centering
\includegraphics[width=0.7\linewidth]{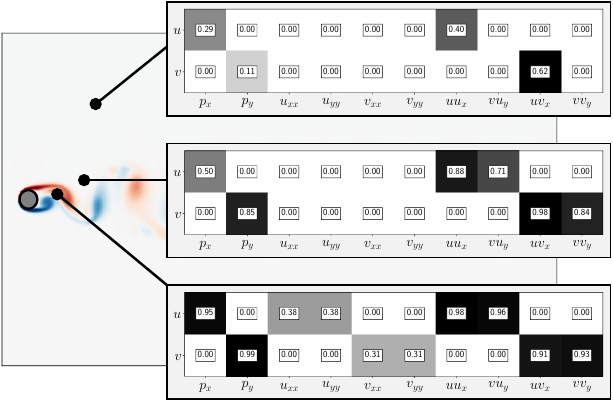}%
\caption{Identified active terms of the library of candidate functions for three random points \cite{costanzo2022}. }
\label{fig: vorticity}
\end{figure}

After the regression, each point is defined by a set of coefficients that indicate the active physical terms that govern the local dynamics. In Figure~\ref{fig: vorticity} active coefficients for three points are shown. The difference between active terms suggests that these three regions might well be governed by different dynamics. 

When applied to a large set of random points, the algorithm would allow the identification of different dynamics present in the domain. However, to obtain some physical insight, the existing dynamics in the domain should be clustered based on the set of active terms calculated for each of the points. Thus, by applying clustering we could in principle regroup the points based on similar dynamical behavior. The most suitable algorithm for the clustering procedure is the Newman network since it does not need an \textit{a priori} knowledge of the number and size of the final clusters allowing to maintain identification process user independent. \\

\subsection{Clustering}
\label{sec:clustering}

The second step of the BIA algorithm consists of applying a clustering algorithm on grid points, where each of the points is described with equation-based feature space defined earlier. For this purpose, Newman's spectral algorithm for community detection in a network \cite{newman2006} is used. As previously mentioned the number of clusters is not defined {\it{a priori}}, instead it is the result of a maximization procedure applied to the network's modularity $Q$.

For this purpose, the Euclidean distance matrix $\Delta$ of the dataset is computed on equation-based feature space:
\begin{equation}
    \Delta_{ij} = \lVert \mathbf{Y}_i - \mathbf{Y}_j \rVert^2 \quad \text{for} \quad (\mathbf{Y}_i,\mathbf{Y}_j) \in \mathbf{Y}^2.
\end{equation}

The dataset is then transformed into an undirected network using a binary adjacency matrix $A$ defined as:
\begin{equation}
    A_{ij} = \left\{
    \begin{array}{lcl}
        1 &\phantom{1234}& \mbox{if} ~ \Delta_{ij}<\epsilon,  \\
        0 && \mbox{otherwise.}
    \end{array}
\right.
\end{equation}

In an undirected network, two points are connected with an edge if their Euclidean distance $\Delta_{ij}$ is below a certain threshold $\epsilon$. 
This threshold is typically selected as a fraction of the average value of the distance matrix $\Delta$. 

Once the network is defined, the split is done progressively into two communities until the modularity $Q$ is maximized. Modularity measures how many edges fall within a community compared to what would be expected in a random network. In other words, let $k_i$ represent the number of edges connected to data point $i$, and $m$ be the total number of connections in the network. Then, the probability of having an edge between $i$ and $j$ in the random reference network is $k_ik_j/m$. Hence, the modularity $Q$ is defined as:

\begin{equation}
    Q = \frac{1}{m}\sum_{ij}\left( A_{ij}-\frac{k_ik_j}{m}\right)\delta_{c_i,c_j},
\end{equation}
where $\delta_{c_i,c_j}$ is the Kronecker delta which is $\delta_{c_i,c_j}=1$ only when $i$ and $j$ belong to the same community, otherwise it is null. The previous equation can be reformulated by defining a vector $\textbf{s}$, where $s_i=1$ if vertex $i$ is in the first group, and $s_i=-1$ otherwise. So $Q$ can be defined as:
\begin{equation}
    Q = \frac{1}{2m}\sum_{ij}\left( A_{ij}-\frac{k_ik_j}{m}\right)(s_is_j+1)=\frac{1}{2m}\mathbf{s^T} B \mathbf{s},
\end{equation}
where the modularity matrix $B$ is given as:
\begin{equation}
    B_{ij} = A_{ij}-\frac{k_ik_j}{m}.
\end{equation}
Since the graph is undirected, the modularity matrix $B$ is symmetric and the modularity $Q$ represents a Rayleigh quotient for matrix $B$. To maximize $Q$, we need to choose a vector $s$ to align with the largest eigenvector  $\textbf{v}$ of $B$, that can be achieved by assigning $s_i=1$ if $v_i>0$ and $s_i=-1$ if $v_i<0$. 

To divide the graph into more than two communities, this process is repeated until further division no longer increases the modularity of each subgraph. A detailed explanation of this algorithm is available in \cite{newman2006}, and also presented in Algorithm~\ref{algo: newman}.

\begin{algorithm}[H]
\label{algo: newman}
\caption{Budget Identification Algorithm (BIA) - clustering}
\KwIn{Data matrix $Y \in \mathbb{R}^{points \times n}$ with $n$ number of features, threshold $\epsilon$}
\KwOut{Number of detected clusters $nc$, cluster labels $Ci \in \mathbb{R}^{n}$ for $Y$}

\Begin{
    Compute the Euclidean distance matrix $\Delta$ for all points $(Y_i, Y_j) \in Y$\;
    Construct adjacency matrix $A$: \;
    \For {each $(i,j)$ pair in $\Delta$}{
        \eIf{$\Delta_{ij} < \epsilon$}{
            $A_{ij} \gets 1$ 
        }{
            $A_{ij} \gets 0$
        }
    }
    Remove self loops $A = A - \text{diag}(A)$ and symmetrize $A \gets \frac{A + A^\top}{2}$\;

    Initialize modularity matrix $B$\;
    Compute degree vectors $K_i$, $K_o$ and total edge weight $m$\;
    Set $B = A - \frac{K_o \times K_i}{m}$\;

    \While {modularity $Q_i > Q_{i-1}$ }{
        Perform eigendecomposition on $B$ and update $Ci$ based on the sign of eigenvector\;
        Recompute $B$ for further partitioning\;
        }
    \Return cluster labels $Ci$, and number of clusters $nc$\;
}
\end{algorithm}

\subsubsection{Clustering applied to flow around a cylinder}

\begin{figure}[ht]
    \centering
    \subfloat[][{$d = 5$ and $n _ \mathrm{ens} = 4$.}]{\includegraphics[width=0.9\linewidth]{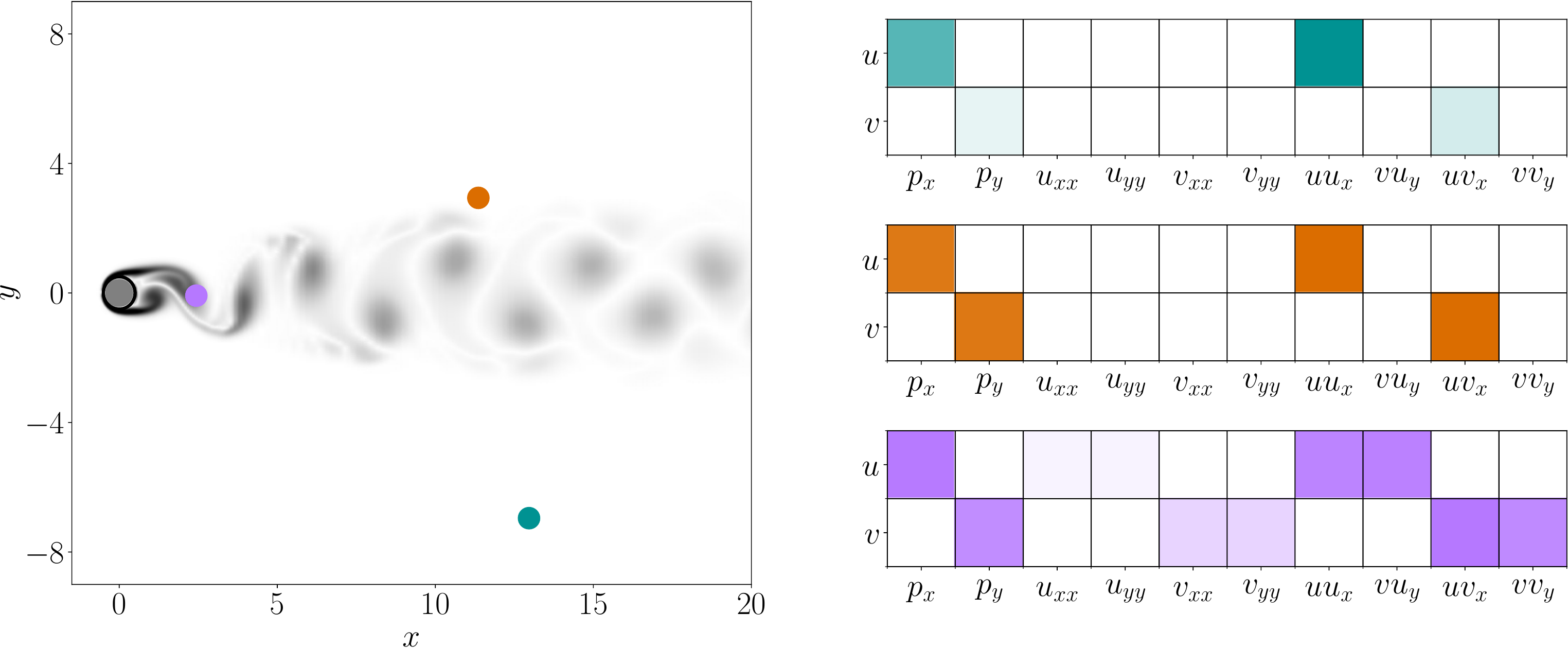}} \\
    \subfloat[][{$d = 5$ and $n _ \mathrm{ens} = 3$.}]{\includegraphics[width=0.31\linewidth]{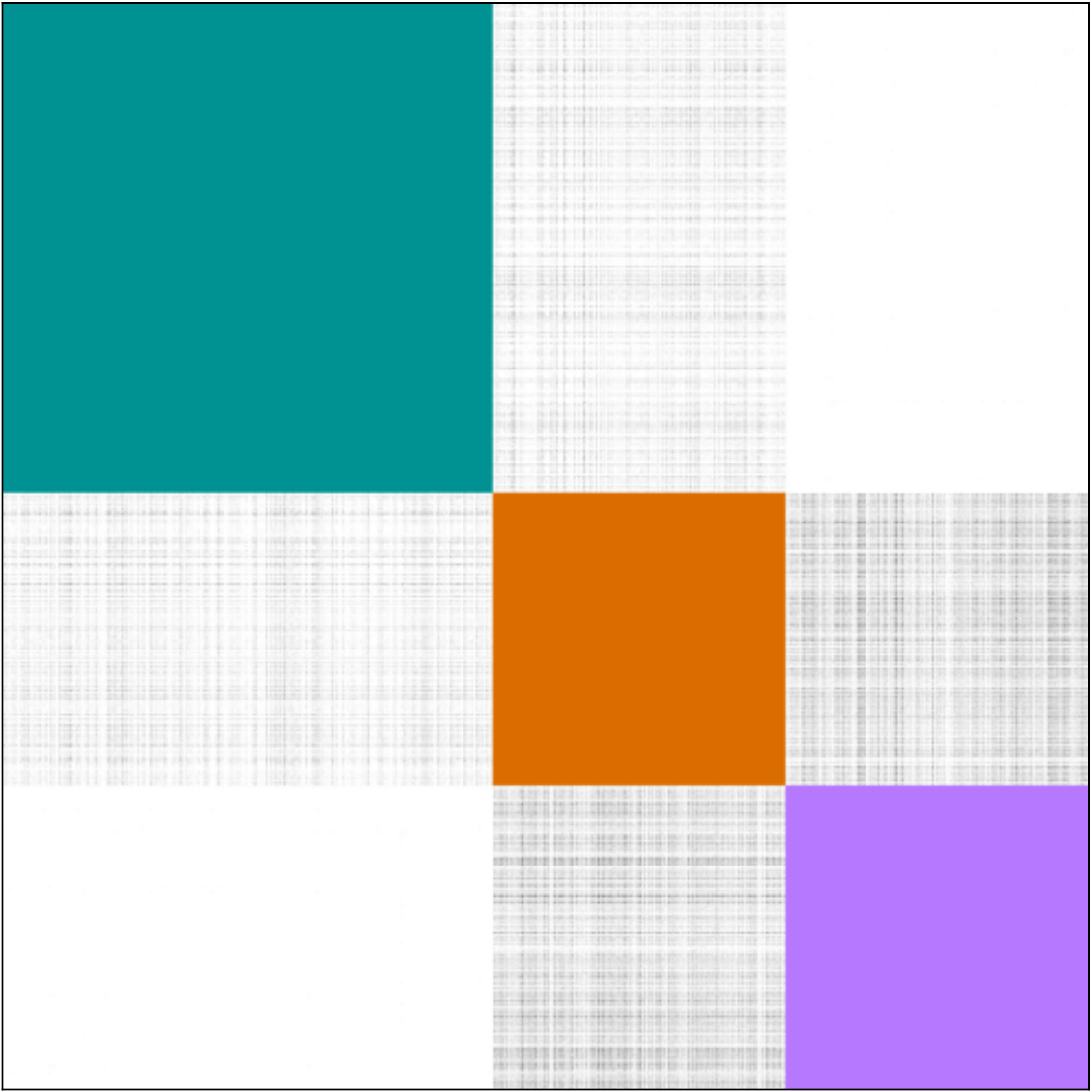}} 
    \vspace{0.3in}
    \subfloat[][{$d = 5$ and $n _ \mathrm{ens} = 1$.}]{\includegraphics[width=0.51\linewidth]{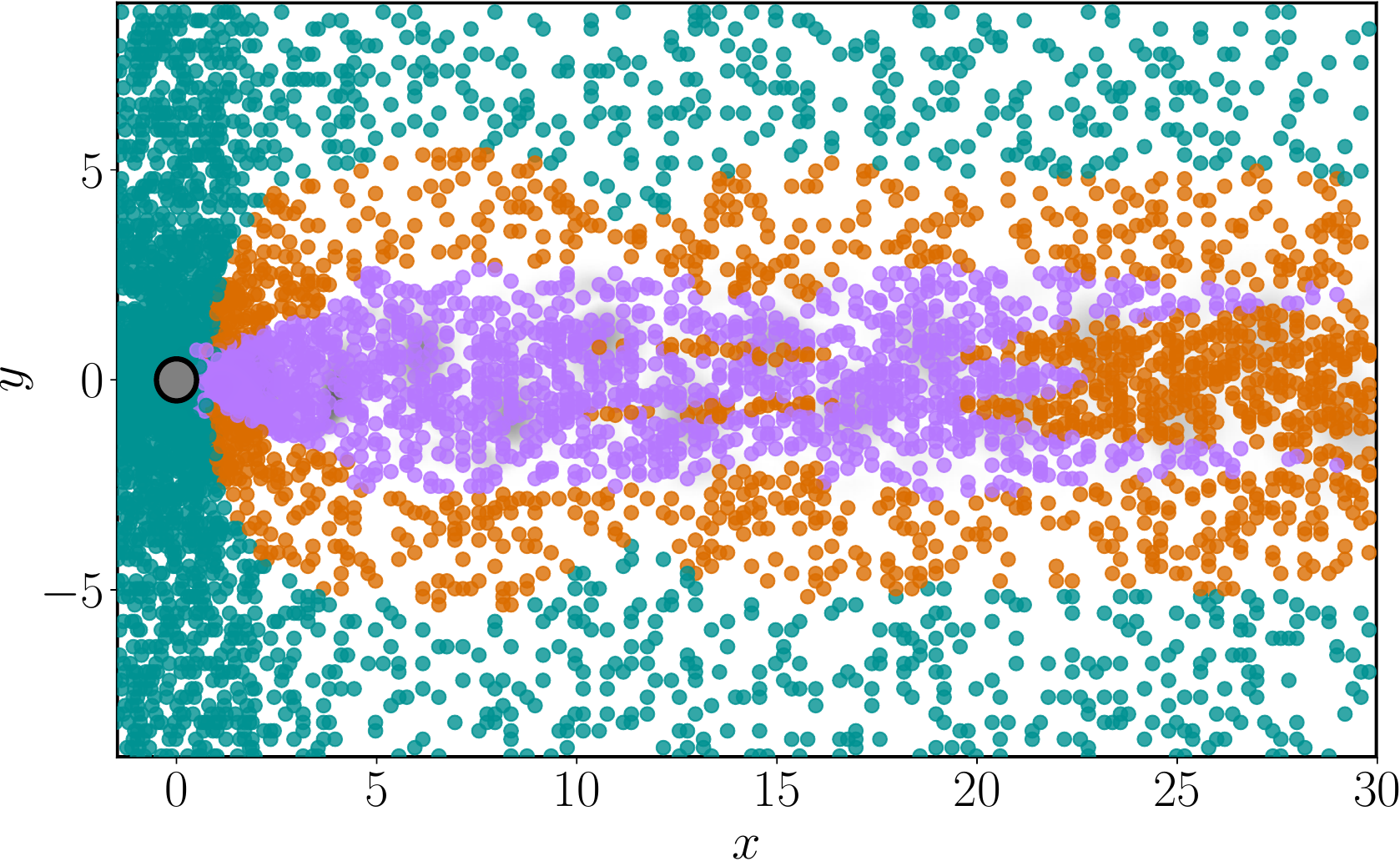}} 
    \caption{Dynamics identification on a two dimensional flow around a cylinder at $Re=200$. (a) the selected points are divided into three main communities sharing same dynamics; (b) rearranged adjacency matrix showing correlations between communities; (c) example of active coefficients identified for points belonging to different communities, in the figure the values are normalized for each point \cite{costanzo2022}.}
    \label{fig: cluster LN}
\end{figure}

\begin{table}[tb]
\begin{center}
\begin{tabularx}{0.9\textwidth} { 
   >{\centering\arraybackslash}X 
  | >{\centering\arraybackslash}X 
  | >{\centering\arraybackslash}X 
  | >{\centering\arraybackslash}X 
  | >{\centering\arraybackslash}X 
  | >{\centering\arraybackslash}X 
  | >{\centering\arraybackslash}X | }
\cline{2-7}
  &  \multicolumn{2}{|c|}{\color{BlueGreen}Cluster 1} & \multicolumn{2}{|c|}{\color{YellowOrange}Cluster 2} & \multicolumn{2}{|c|}{\color{Orchid}Cluster 3} \\ \cline{2-7}
  &$u$ & $v$ &$u$ &$v$ & $u$ & $v$ \\  \hline%
  \multicolumn{1}{|c|}{} & & & & & & \\[-1em]
 \multicolumn{1}{|c|}{$p_{x}$} &$0.013$& &$0.89$ & &$1.015$ &\\ [1mm]\hline
  \multicolumn{1}{|c|}{}& & & & & & \\[-1em]
  \multicolumn{1}{|c|}{$p_{y}$}& &$0.002$ & &$0.89$ & &$0.879$\\[1mm] \hline 
   \multicolumn{1}{|c|}{}& & & & & &\\[-1em]
    \multicolumn{1}{|c|}{$\frac{1}{Re}u_{xx}$}& & & & &$0.135$ &\\ [1mm]\hline 
    \multicolumn{1}{|c|}{}& & & & & & \\[-1em]
   \multicolumn{1}{|c|}{$\frac{1}{Re}u_{yy}$}& & & & &$0.135$ &\\[1mm]\hline 
  \multicolumn{1}{|c|}{} & & & & & &\\[-1em]
  \multicolumn{1}{|c|}{$\frac{1}{Re}v_{xx}$}& & & & & &$0.361$\\ [1mm]\hline 
   \multicolumn{1}{|c|}{}& & & & & &\\[-1em]
  \multicolumn{1}{|c|}{$\frac{1}{Re}v_{yy}$}& & & & & &$0.361$\\[1mm] \hline
   \multicolumn{1}{|c|}{}& & & & & & \\[-1em]
   \multicolumn{1}{|c|}{$uu_{x}$}&$0.02$ & &$0.97$ & &$0.946$ &\\ [1mm]\hline
   \multicolumn{1}{|c|}{}& & & & & & \\[-1em]
  \multicolumn{1}{|c|}{$vu_{y}$}& & & & &$0.946$ &\\ [1mm]\hline
  \multicolumn{1}{|c|}{} & & & & & & \\[-1em]
   \multicolumn{1}{|c|}{$uv_{x}$}& &$0.004$ & &$0.97$ & &$1.035$\\[1mm] \hline
   \multicolumn{1}{|c|}{}& & & & & & \\[-1em]
  \multicolumn{1}{|c|}{$vv_{y}$}& & & & & &$0.899$\\ [1mm]\hline
\end{tabularx}
\end{center}
\caption{Numerical values of the identified coefficients for points belonging to different communities \cite{costanzo2022}.}
\label{tab: coeff}
\end{table}

In the example of Figure~\ref{fig: vorticity and points} we show the coefficients only for three points. We proceed to apply regression on more points to proceed with clustering of flow dynamics.
After the regression, each point is transformed into the coefficient subspace  $\boldsymbol{\Xi}^{(j)}$ and is set to be a node of the Newman network. Thus, the edges are set according to the Euclidean distance in the coefficient subspace and the selected threshold. In this example, the threshold is set to $10\%$ of the maximum distance among all the points included in the procedure.

With the Newman network we could identify three communities shown in the Figure~\ref{fig: cluster LN} (c). Three clusters are: a free stream region (blue-green) where the flow is not perturbed by the presence of the cylinder, a central region behind the cylinder (violet) where the von Karman vortex is evolving, and an intermediate layer (orange) between the two. As we move away from the cylinder in the positive $x$-direction, violet and orange clusters start to merge, due to the reduced intensity of the vorticity field. 

The rearranged adjacency matrix of the clustering is shown in Figure~\ref{fig: cluster LN} (b), where the three detected clusters are colored with its corresponding color. The grey areas in the adjacency matrix show that the identified clusters present correlations and are not independent i.e. the dynamics governing each cluster can affect the other. 

For understanding the physical meaning of the clusters we now focus on analyzing the results represented in Figure~\ref{fig: cluster LN} (a), where we show the active coefficients found through linear regression for three points from different clusters. It should be noticed that each set of coefficients has been scaled based on its maximum value. In the violet cluster behind the cylinder, all the terms on the right-hand side of the Navier-Stokes equation are active. However, in the orange cluster, as we move further from the wake, the effect of viscous forces gradually decreases until they become inactive. In the free stream, green-blue cluster, the pressure and inertial forces are still present, but their values are much smaller compared to those in the other clusters. For reference, the numerical values of the coefficients for these points are shown in Table~\ref{tab: coeff}. Based on these results, we can conclude that the clusters have strong physical interpretations, and the BIA algorithm is effective.\\

\section{Dynamic-BIA -- Lagrangian framework}
\label{sec:dyn-bia}
Up to this point, the algorithm has been used in the Eulerian framework, which means that each point remains stationary in space as the flow evolves. While this is not a problem for stationary systems, it may be more informative to let each point evolve in space and time as the regression is performed, moving towards a Lagrangian framework. In this section we present a dynamic version of the BIA algorithm which accounts for the local movement of each regression point and therefore the resulting clusters.   

In the dynamic framework, each point represents a particle in the flow. In order to account for the movement of each particle, its local velocity is stored in memory. For example, starting from timestep 1, we will store the velocity in the $x$ and $y$ directions of each particle/point, which will determine where the particle/point will move in the next timestep. We then construct the trajectory of each particle by summing the displacements (velocity updated each timestep and multiplied by time) starting from timestep 1. In step 2 the variables such as $u$,$v$,$w$,$\rho$ associated with the new position of the particle based on its displacement are then used to build each term of the equation. The regression and clustering step of the BIA algorithm is then applied to each particle trajectory in the flow. 

After applying the dynamic-BIA algorithm to each particle and performing Newman clustering on the equation-based feature space, we can reconstruct the movement of the clusters. This is possible because each particle's trajectory has been stored during the regression step, and once the particle is assigned to one of Newman's clusters, this trajectory is given a cluster label.

\subsection{Dynamic-BIA applied to flow around a cylinder}
\begin{figure}[tb]
\centering
\includegraphics[width=1\linewidth]{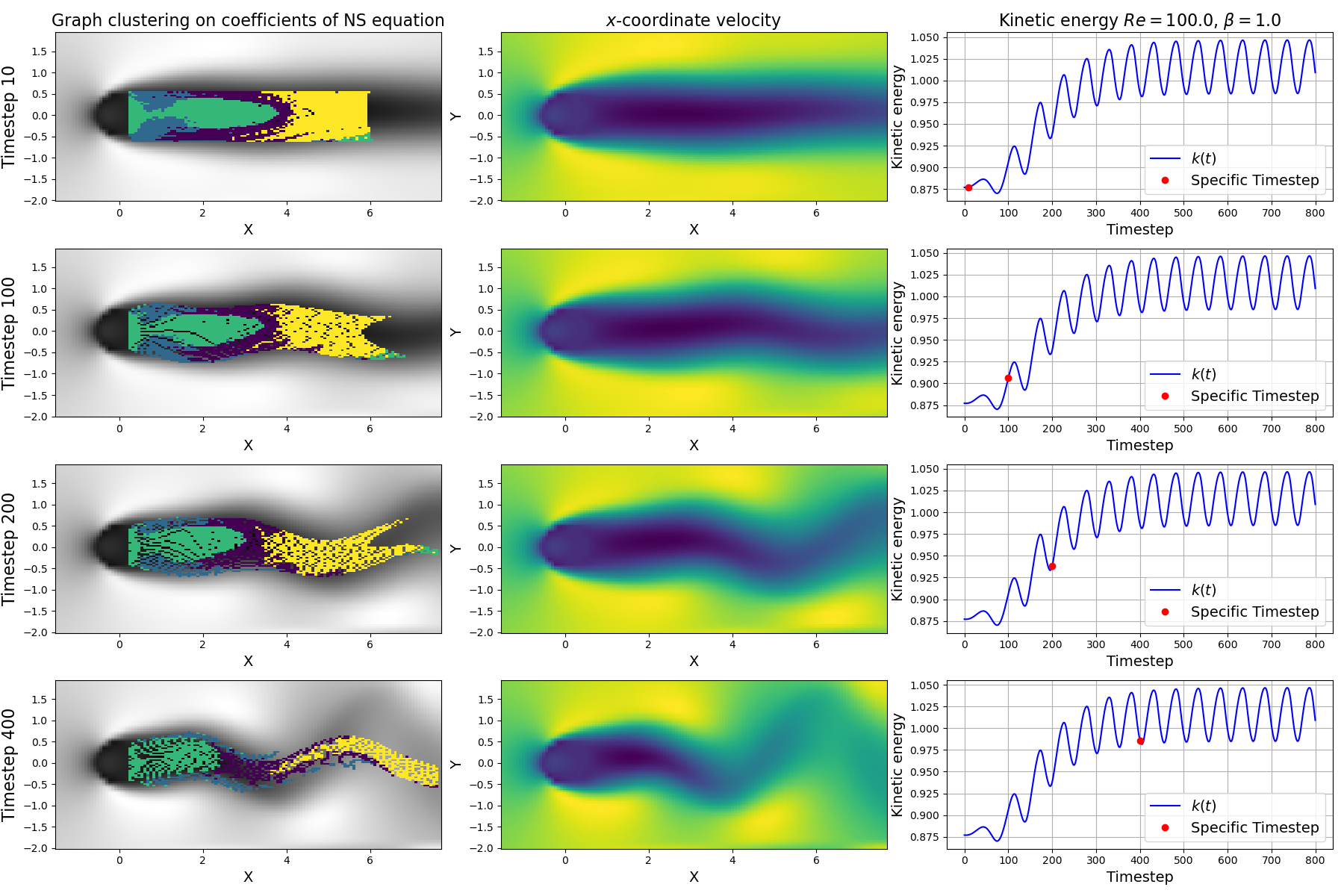}%
\caption{Application of BIA algorithm for NS equation equation on flow around the cylinder: Newman clustering of the coefficients of the NS equation
(4 clusters, yellow/purple/green/blue) (left), representation of the $x$-coordinate velocity (center), evolution of the total kinetic
energy of the flow over time with red dot marking the time instant for each row (right).}
\label{fig:NS_BIA_cluster}
\end{figure}

The result of the dynamic BIA is first assessed using the same data of a flow around a cylinder as in section~\ref{sec:BIA}. The two main differences are that firstly, in the previous section, only the data coinciding with the flow at the stable limit cycle, where the von-Karman streak appears, is considered. This region can be seen in the last column of Figure~\ref{fig:NS_BIA_cluster}, representing the snapshots beyond $t=400$. By removing the transient region, the application of BIA in the Eulerian framework can be supported. However, now that the Largrangian framework is considered, we have included the transient regime in the analysis. The flow starts from the unstable initial state and evolves to the stable limit cycle where the shedding is established. The aim of including the transient regime is to test whether the clusters are able to adapt to the changing dynamics of the flow. The second difference is the change in the sampling points. While in the previous section the whole domain is sampled, in the Lagrangian version, only the region behind the wake is considered. This is mainly because the points initialised in the free-stream quickly move out of the domain without contributing to the dynamics due to the convective nature of the flow. To keep the sampling space filling, they must then be re-initialised, making the algorithm less efficient.

\begin{figure}[ht]
    \centering
    \includegraphics[width=1\linewidth]{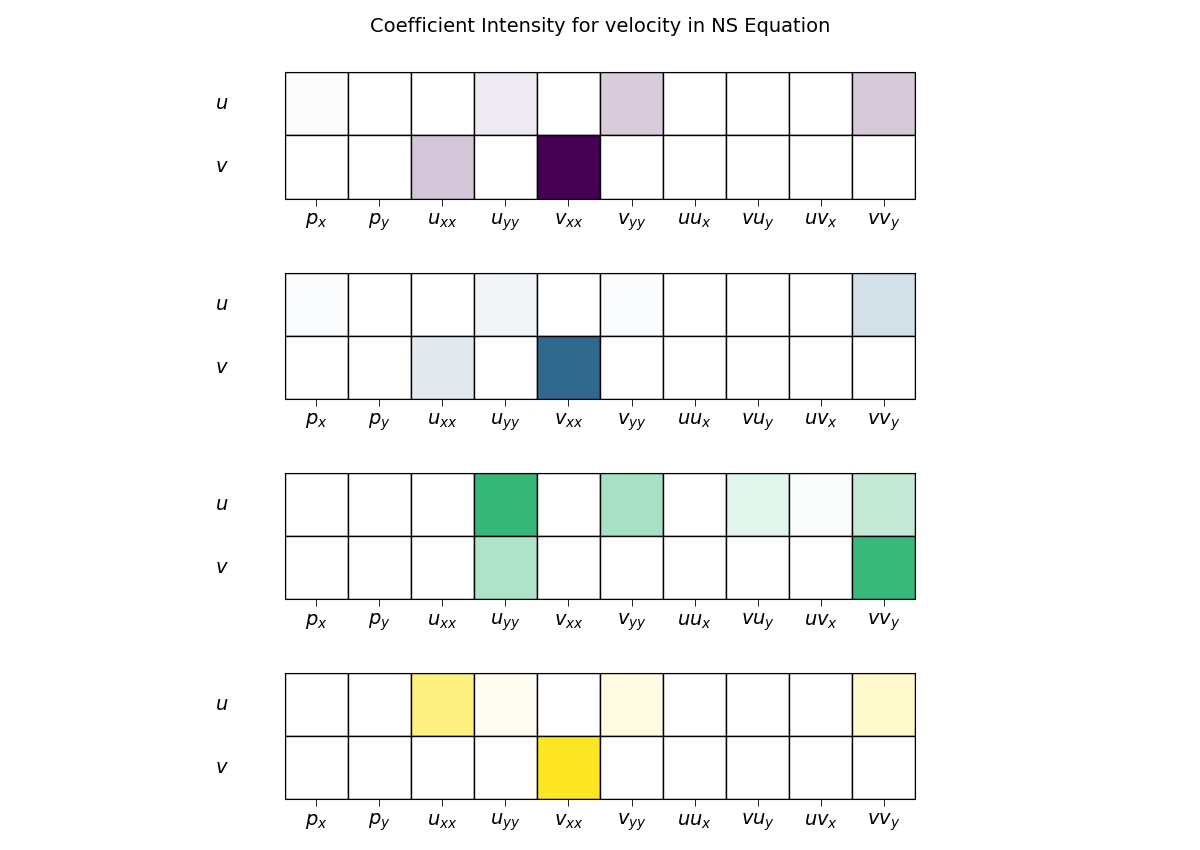}%
    \caption{Application of BIA algorithm for NS equation:  Representation of active coefficients for purple/blue/green/yellow clusters from Figure~\ref{fig:NS_BIA_cluster} for NS equation. Each cluster is assigned a colormap of corresponding color representing the level of contribution of each term; darker tones correspond to maximum ($max=1$) and lighter tones to minimum ($min=0$) contributions.
    }
    \label{fig:ns_intensity}
\end{figure}

Figure~\ref{fig:NS_BIA_cluster} shows the performance of the Dynamic-BIA algorithm. The first column shows the evolution of the clusters at different time steps. The second column juxtaposes the flow at each instance, using the contours of the streamwise velocity component, and the last column shows the evolution of the total kinetic energy of the flow over time, where each red dot marks the location of the time instant used to extract the plots of the first and second columns. The kinetic energy is included as a representative of the flow regime. The transient regime is highlighted by the increase in kinetic energy and the limit cycle regime can also be identified by the region where the kinetic energy oscillates around a stationary mean value; i.e. beyond $t = 400$. Figure~\ref{fig:ns_intensity} again shows the contribution of each term in the equation depending on the cluster to which it belongs. Similar to the clusters identified by BIA in the Eulerian framework, the green cluster of the near wake region has the most terms in the equation active.

Focusing on $t = 10$, this regime corresponds to the base-state solution and, as expected, the clusters identify a large wake (separation bubble) behind a cylinder in a symmetric form. The separated region then transitions to a yellow cluster regime, which identifies the wake behind the separation bubble, corresponding to the region where the flow has reattached. These flow structures agree well with what is known from the base-state solution of cylinder flow at sufficiently low Reynolds numbers. As time progresses into the transition regime, the bubble behind the cylinder becomes asymmetric and the constructed cluster (green in the figure) starts to oscillate at a frequency matching that of the flow behind the cylinder. This confirms that the identified clusters are able to follow the dynamics of the specific region to which they are assigned. This also highlights the importance of the Lagrangian framework. As the region with the identified dynamics changes spatially, Dynamic BIA is able to adapt the movement of each cluster. While the BIA algorithm of the previous section assigns the whole region behind the cylinder to one cluster, the Lagrangian framework is able to distinguish the near separation from the following wake behind a cylinder, and reconstruct the movement of this cluster.  Finally, the clusters behind the cylinder settle into a smaller bubble when the limit cycle is established and can follow the correct oscillations according to the Strouhal number of the shedding.

The simplified cylinder case provides a test bed to validate and assess the performance of the algorithm. In the next section, the flexibility of the approach is tested by considering a turbulent flow. 

\section{Application to turbulent stratified flows}
\label{sec:turbulence}

Stably stratified turbulence (SST) occurs when there is a stabilising density gradient due to variations in temperature or species concentration. It is a model flow for a wide range of environmental and engineering flows. It is also an interesting test case for current methodology, as it involves a three-way competition between inertial, buoyant and viscous forces, often resulting in a constantly changing amalgamation of dynamically distinct regions \citep{portwood16}. Understanding, and possibly clustering, these dynamics can give an indication of how turbulence evolves, how kinetic energy is distributed, and possibly how the density staircases so important for mixing heat in the ocean are formed \citep{caulfield2021}.\\

\begin{figure}[tbh]
\centering
\includegraphics[width=1\linewidth]{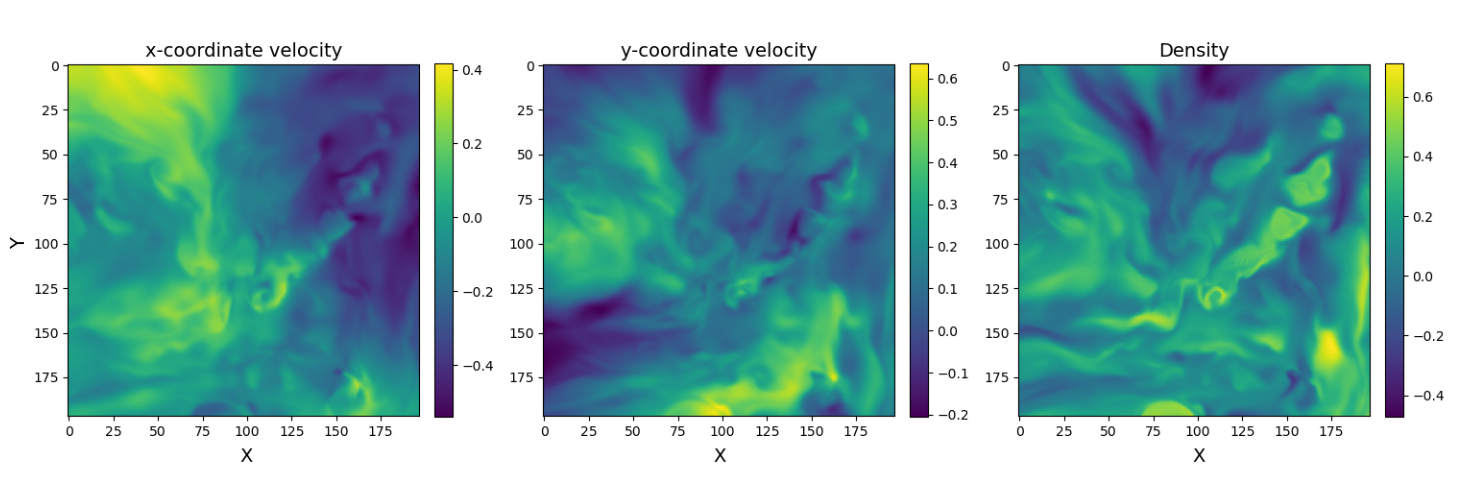}%
\caption{Representation of horizontal, vertical velocities and density for $t$=25.2 [s] of stably stratified turbulence simulation $Re=3200$ and $Fr=32$.}
\label{fig: data_set_stratified_flow}
\end{figure}

Simulations of SST are often performed assuming the Boussinesq approximation for the density, so that the kinetic and internal energies are decoupled, and the kinetic and potential energies are coupled only through the buoyancy flux.  This leads us to consider for our analyses the turbulent kinetic energy (TKE) and the turbulent buoyancy variance (TBV), which are defined in section~\ref{sec:governing_equation}. These two scalar equations are among the simplest-but-not-simpler for providing insight into the turbulence and associated energy fluxes. By comparing the clusters identified from each equation, we can better understand the turbulence dynamics and also draw some conclusions about the most appropriate equation to represent the dynamics of such flows.\\

The data we use here are extracted from a direct numerical simulation initialised with Taylor-Green \citep{taylor37} vortices plus low lever noise \citep{riley03,hebert06b}. The ambient stable density gradient is constant in time and aligned with gravity. Some recent developments in understanding this flow are presented in \citep{petropoulos24c}. Characteristics that make it particularly interesting for testing the current methodology is this: it starts out laminar, but with the buoynacy force not balanced by the pressure gradient so that, immediately, energy begins transferring between kinetic and potential energy.  Because of the low level noise, the flow is unstable and becomes turbulent by about dimensionless time $t=5$[s]. By $t=15$[s], the turbulence intensity peaks and then decays as the flow relaminarises.\\

The particular case we consider has Reynolds number $Re=3200$ and Froude number $Fr=32$, both based on the initial Taylor-Green vortex velocity and length scales (see Figure~\ref{fig: data_set_stratified_flow}).  Despite the fairly large value of $Fr$, the buoyancy force is of first-order importance throughout the simulation, and the flow is becomes very strongly stratified as it decays \citep{hebert06b}.  The simulation variables are the fluctuating density $\rho$ and the fluctuating velocities $[u,v,w]$ in the $x$,$y$, and $z$ directions, respectively.  Each variable is a four-dimensional tensor of dimensions $512 \times 512 \times 256 \times$ timesteps, where the first three dimensions are spatial.\\

\subsection{Governing equations}
\label{sec:governing_equation}
As mentioned in section~\ref{sec:BIA}, before applying BIA, it is necessary to first create a specialised candidate library containing terms from the selected equation. In our case, we will create a library for two types of equations, the turbulent kinetic energy (TKE) and turbulent buoyancy variance (TBV) equations, and independently evaluate how clustering performs for each. Both TKE and TBV equations are extracted from \cite{caulfield2021} and are derived under the following assumptions: (i) the gravitational force acts in the $z$-direction, (ii) the flow is incompressible, i.e. the velocity field satisfies the divergence-free condition, and (iii) the Boussinesq approximation holds by considering density fluctuations only for the buoyancy force term. Finally, Reynolds decomposition is applied to each variable in the equations as follows:
\begin{equation}
    u = U + u', \quad \quad \quad  v = V + v', \quad \quad \quad w = W + w', \quad \quad \quad \rho = \rho + \rho'
\end{equation}
where \(\ang{\cdot}\) is the ensemble average that can be considered as spatial or temporal average and  \( u' \), \( v' \), \( w' \), \( \rho' \) are the fluctuating components and  \( U \), \( V \), \( W \), \( \rho \) are the mean components of the variables. As the simulation is isotropic, the mean components in the Reynolds decomposition are zero.  

Following the above mentioned assumptions, the TKE equation for which the family library is further constructed is defined as:
\begin{equation}
\begin{aligned}
    \overbrace{\pder{}{t} \left( \frac{1}{2} \ang{u_i' u_i'} \right)}^{\mathcal{K}'} &= 
    \overbrace{- 2 \nu \ang{s_{ij}' s_{ij}'}}^{\mathcal{E}'} - 
    \overbrace{\frac{g}{\rho_r} \ang{\rho' w'}}^{\mathcal{B}} +
    \overbrace{\left( -\ang{u_i' u_j'} S_{ij} \right)}^{\mathcal{P}} \\
    &\quad - \underbrace{\pder{}{x_i} \left[ \ang{p' u_i'} 
    + \ang{\left( u_i' + U_i \right) \frac{u_j' u_j'}{2}} 
    - \nu \left( \pder{}{x_i} \ang{\frac{u_j' u_j'}{2}} 
    + \pder{}{x_j} \ang{u_i' u_i'} \right) \right]}_{\nabla \cdot J}
\end{aligned}
\label{eq:tke}
\end{equation}

\noindent where \( p'\) is the pressure fluctuations, \( g \) gravity acceleration, \( \nu \) the kinematic viscosity, $s_{ij}' = \frac{1}{2}\left(\frac{\partial u_i'}{\partial x_j} + \frac{\partial u_j'}{\partial x_i}\right)$ and $S_{ij} = \frac{1}{2}\left(\frac{\partial U_i}{\partial x_j} + \frac{\partial U_j}{\partial x_i}\right)$ are respectively the perturbation strain-rate tensors and the mean strain-rate tensors. 

Following the same procedure for the Turbulent Buoyancy Variance, the candidate library will include the terms defined by the following equation,

\begin{equation}
\begin{aligned}
    \frac{1}{N^2} \pder{}{t} \ang{\frac{b^2}{2}} = 
    \underbrace{\frac{g}{\rho_r} \ang{\rho' w'}}_{\mathcal{B}} - 
    \underbrace{\frac{\kappa}{N^2} \ang{\pder{b}{x_i} \pder{b}{x_i}}}_{\chi} -
    \nabla \cdot \mathbf{J}_{\rho}   \\
    \nabla \cdot \mathbf{J}_{\rho} = 
    \sum_{i=1}^{3} \frac{1}{N^2} \pder{}{x_i} \left[ U_i \ang{\frac{b^2}{2}} + u'_i b' - \kappa b \pder{b}{x_i} \right]  \hspace{1cm} N^2(z,t) \equiv - \frac{g}{\rho_r} \pder{}{z} \ang{\rho} (z,t)
\end{aligned}
\label{eq: tbv}
\end{equation}
where $b$ is buoyancy defined as $b = \rho' \frac{g}{\rho_r}$, $\kappa$ the molecular diffusivity and $N$ buoyancy frequency. Note if $\rho$ is a linear function of $z$ alone, and hence $N$ is a constant, the TBV equation is equivalent to a (turbulent) available potential energy equation, making more intuitive the exchange between the kinetic and (available) potential energy reservoirs via $\mathcal{B}$. For more detail regarding the derivation of each equation please refer to \cite{caulfield2021}.

Each term from Eqs.~\ref{eq:tke}, \&~\ref{eq: tbv} should be expanded for \( i = 1, 2, 3 = x, y, z \) considering that \( u_x = u_1 = u \), \( u_y = u_2 = v \), and \( u_z = u_3 = w \). The fully expanded equations with the explanation of the physical meaning of each term can be found in~\ref{TKE_appendix_development}, and~\ref{TBV_appendix_development} respectively.


\subsection{Data preparation}
Since the data does not include all the necessary variables to construct a complete family library for TBV and TKE equations, several simplifications are made. Firstly, due to the absence of pressure in the data, we implicitly impose the divergence-free condition by projecting the field velocity onto its divergence-free component, thus eliminating the pressure term from the TKE equation. Secondly, due to the isotropic nature of the flow, we have removed all terms including the mean velocity component from the library. As mentioned above, the initial cyclostrophic imbalance condition tends to create a $z$-axis mean flow, which unfortunately cannot be calculated or considered in the family library of either equation. The solution to this is to select only the last time steps of the turbulence, i.e. $t=27$ [s] to $t=30$ [s], as the vertical velocity and shear created by the initial cyclostophic imbalance would have already transformed into non-isotropic turbulence and the mean flow would have completely decayed with kinetic energy. Another advantage of considering only the last time steps is that the most non-isotropic large turbulent kinetic energy scales are transformed into smaller ones closer to the isotropic regime. As this type of regime is more uniform, the predicted equation for turbulence with the BIA algorithm can be made with greater accuracy. As the flow is not statistically stationary in the $z$-direction, we also applied BIA to two-dimensional $x$-$y$ slices in the dataset. 

It should be noted that in the Eqs.~\ref{eq:tke}, \& \ref{eq: tbv} the ensemble average operand appears in all of the components. This operand can be approximated as a spatial or temporal average. For our case of study, we proceed with spatial averaging and more concretely use Gaussian kernel weighted average defined as,
\begin{equation}
G(z, z_i, \sigma) = \exp\left(-\frac{(z - z_i)^2}{2\sigma^2}\right),
\end{equation}
where \( z \) and \( z_i \) are the input values and \( \sigma \) is the standard deviation of the Gaussian kernel.

\subsection{Adaptation of the algorithm - point-wise to patch-wise}
\begin{figure}[H]
\centering
\includegraphics[width=0.9\linewidth]{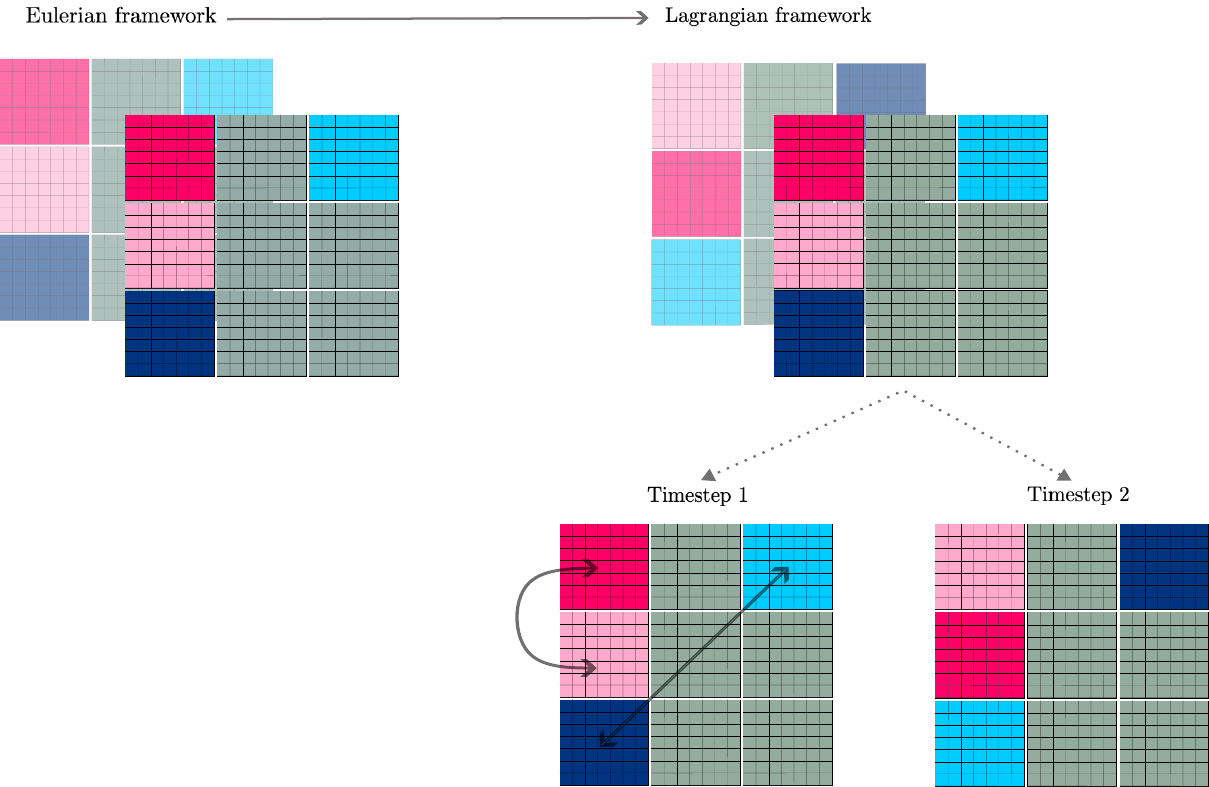}%
\caption{Representation of Lagrangian framework data pre-processing. Considering two timesteps of a spatial grid $3 \times 3$ where each point is characterized with $7 \times 7$ patches. In the Eulerian framework both timesteps maintain the same patch ordering, while in Lagrangian framework timestep 2 is rearranged based on velocity data of each patch from timestep 1.}
\label{fig:framework}
\end{figure}

Due to the nature of spatial-averaged equations, a slight modification to the presented algorithm is necessary. Since working within a single point will not be of relevance to this application, a notion of patches is introduced. Following this strategy, each two-dimensional horizontal slice is divided into small subgrids, also called patches. By pre-processing the original grid into patches, each point of the grid is then characterized not only by itself but also by its surroundings. In Figure~\ref{fig:framework} using the Eulerian framework, we show an example which pre-processes of two timesteps of a spatial grid $3 \times 3$, where each point is characterized by a $7 \times 7$ patch. This pre-processing technique consists of moving the fixed $7\times7$ window over the entire grid. Some points on the edge might end up with no patch assigned. Appropriately justified boundary conditions need to be applied to the edge points, which will be the subject of future work. 

In the Eulerian framework, the variables are recorded over time for fixed patches in space, as the coordinate system is static. As shown in Figure~\ref{fig:framework}, both time steps preserve the same order of the patches. To move to the Lagrangian framework, we assume that the initial spatial $3 \times 3$ patch grid does not represent spatial points but particles (we assign each particle a matrix index corresponding to its initial position on the $3 \times 3$ grid). Once this assumption is made, the rest of the algorithm is identical to that presented in section~\ref{sec:dyn-bia}. Starting from
timestep 1, we calculate the average velocity in the x and y directions of each patch. This determines where the patch will move in the next timestep. We then construct the trajectory of each patch by summing the displacements (average velocity updated each timestep and multiplied by
time) starting from timestep 1. Figure~\ref{fig:framework} shows how, in the second timestep, the variables assigned to the patches can be updated based on their displacement in timestep 1.  

\begin{figure}[H]
    \centering
      \subfloat[][{}]{\includegraphics[width=1\linewidth]{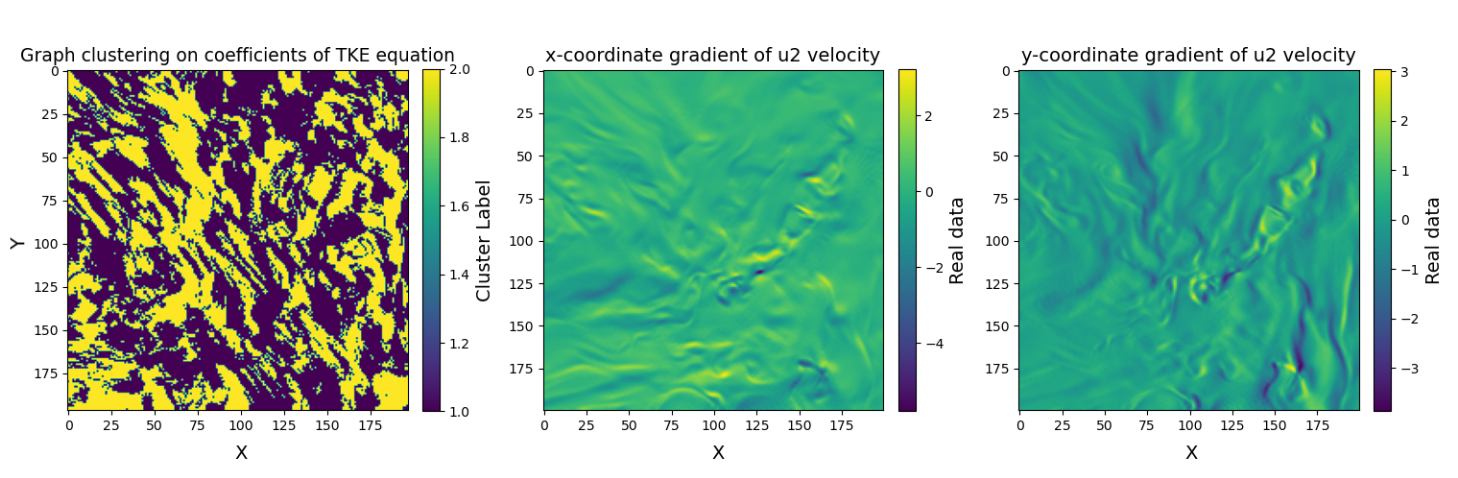}}
    \vspace{0.05in}
    \subfloat[][{}]{\includegraphics[width=1\linewidth]{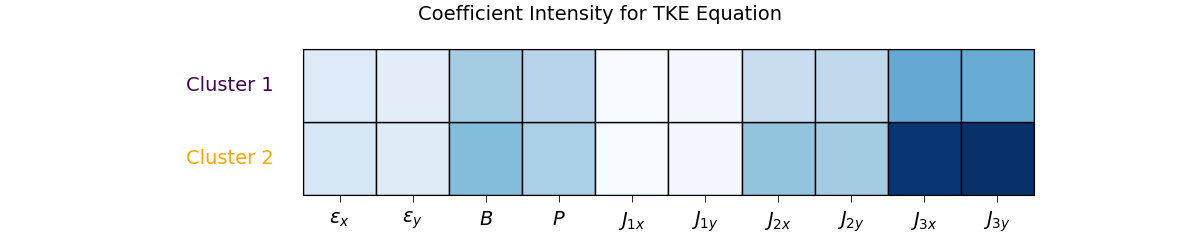}}
    \caption{Application of BIA algorithm for TKE equation: (a) Newman clustering of the coefficients of the TKE equation (2 clusters, yellow/purple) (left), representation of the horizontal velocity (u2) gradient in $x$-coordinate (center), representation of the horizontal velocity (u2) gradient in $y$ -coordinate (right); (b) Representation of active coefficients identified in yellow/purple clusters for TKE equation. The blue colormap represents the level of contribution of each term, where darker tones correspond to maximum ($max=1$) and lighter tones to minimum ($min=0$) contributions.}
    \label{fig: TKE_cludter}
\end{figure}

The use of dynamic clustering in this application would be advantageous because the flow behaviour can be intermittent with varying dynamics over time. If each patch is considered fixed in space, different events will coincide and contribute to the regression and clustering process, making it difficult to decompose the dynamic behaviour of each event. In fact, when the static version of the BIA algorithm is applied (not shown here), no distinct clusters can be identified. 

\begin{figure}[H]
    \centering
      \subfloat[][{}]{\includegraphics[width=1\linewidth]{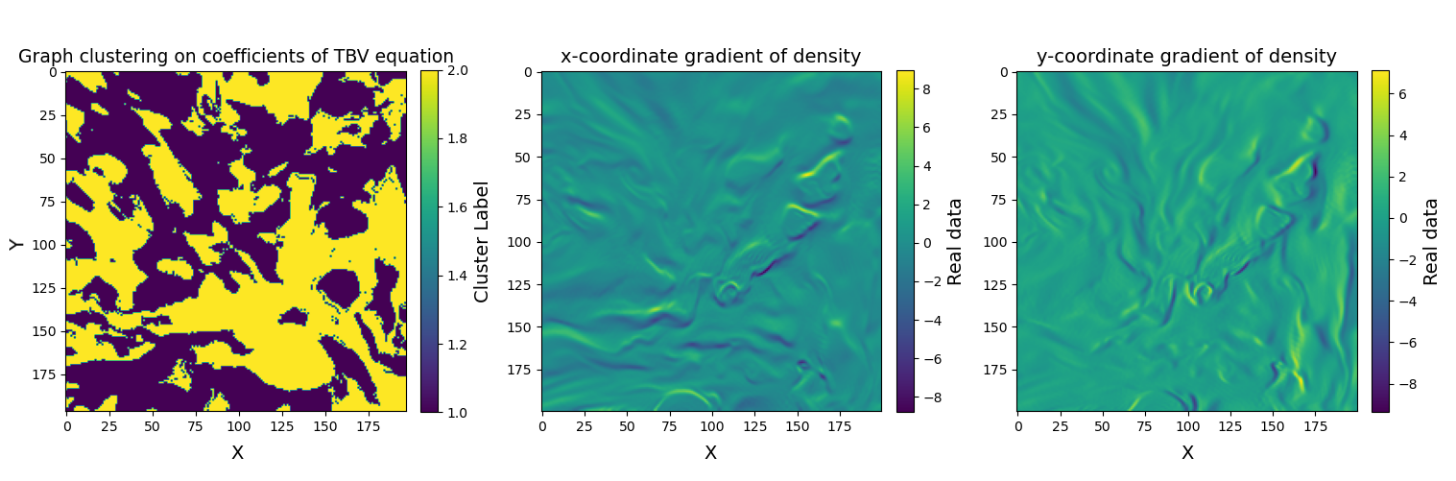}}
    \vspace{-0.005in}
    \subfloat[][{}]{\includegraphics[width=1\linewidth]{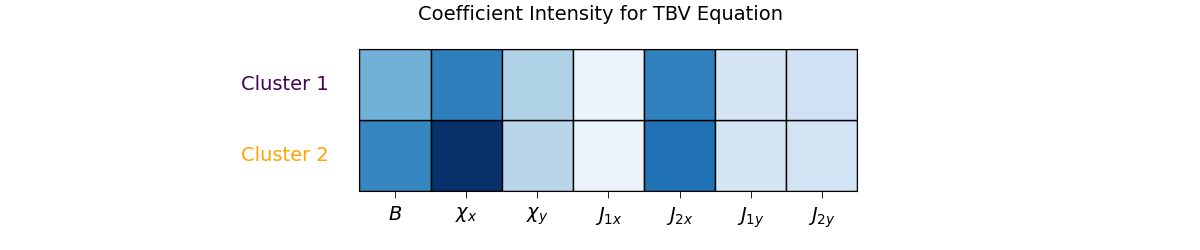}}
    \caption{Application of BIA algorithm for TBV equation: (a) Newman clustering of the coefficients of the TBV equation (2 clusters, yellow/purple) (left), representation of the density gradient in $x$-coordinate (center), representation of the density gradient in $y$ -coordinate (right); (b) Representation of active coefficients identified in yellow/purple clusters for TBV equation. The blue colormap represents the level of contribution of each term, where darker tones correspond to maximum ($max=1$) and lighter tones to minimum ($min=0$) contributions.}
    \label{fig: cluster_tbv}
\end{figure}

\subsection{Results}
\label{sed:results}

The laboratory experiments from \cite{experimental_data_1}, \cite{experimental_data_2} provide an insight into turbulent decaying phenomena in a stratified flow. There is generically a suppression of vertical motion due to buoyancy forces during the time evolution of stratified flows. The vertical motion decay leads to the development of quasi-two-dimensional vortices i.e. horizontal layers of eddies separated by intense shear. These vortices are also called "pancake-like" eddies, as their shape is elongated along the $x-$, $y-$axes and compressed along the $z$-axis. The results shown in Figures~\ref{fig: TKE_cludter}, ~\ref{fig: cluster_tbv} show the same amount of clusters. The yellow cluster represents flow dominated by buoyancy forces, leading to dominant $x$- and $y$-direction transport terms. In contrast, the blue cluster has less pronounced planar transport, suggesting more significant movement in the vertical.

When comparing TBV with TKE clustering in stratified fluid flows, TBV appears to be the better choice. The TBV equation contains most of the terms directly related to the density-driven dynamics governing vertical motion, which is critical to understanding how stratification affects the flow. In contrast, most of the terms in the TKE equation focus on momentum and cannot capture the subtleties of density variations that drive the behavior of stratified layers. TBV clustering therefore provides a clearer picture of how stratification affects flow and offers fewer features for clustering, with all its terms more closely related to density variations, and hence dynamics dominated by buoyancy.

\begin{figure}[htb]
\centering
\includegraphics[width=0.6\linewidth]{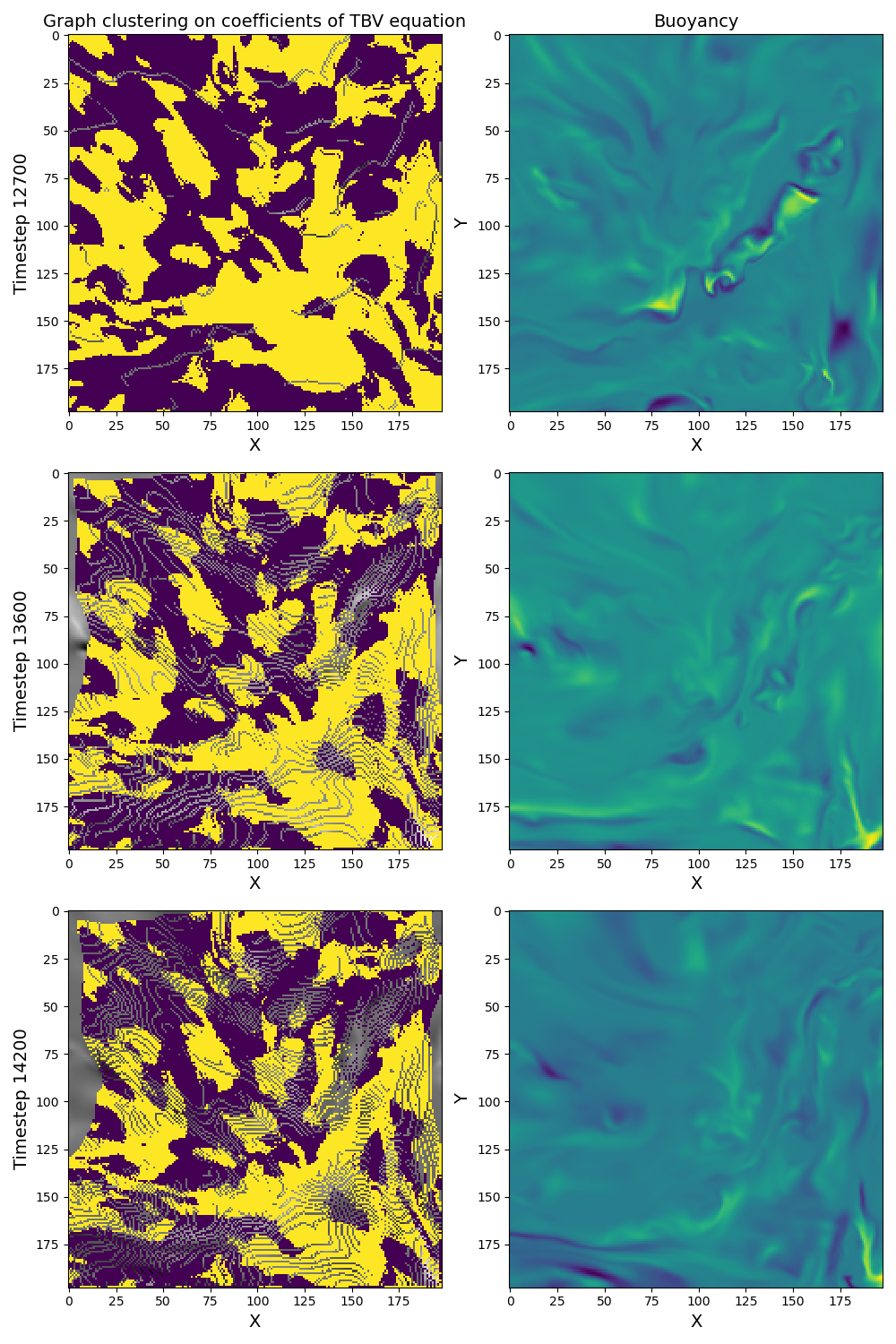}%
\caption{Results of dynamic cluster evolution over three equally separated timesteps (13400, 13800, 14200) corresponding to a time gap of 0.8 [s] between each panel.}
\label{fig: moving_5_timesteps}
\end{figure}

After the cluster identification step, we are able to reconstruct the motion of each cluster by accessing the motion of individual patches stored during the pre-processing step. These dynamic clusters evolve over time based on the velocity of each patch, allowing us to observe how energy budgets move within the flow. In the Figure~\ref{fig: moving_5_timesteps} we have extracted the results of three equally spaced timesteps. We can see a gradual movement of the yellow cluster, with buoyancy dominating the budget, following a diagonal path downwards from the top left corner and progressing towards the bottom right corner. This allows us to analyse how turbulence would evolve. For example, by following the motion of a cluster governed by the buoyancy term, we can predict that turbulence will be suppressed in the direction of motion of that cluster. By tracking in time, we can further optimise computational resources by predicting which region in the next time step should be prioritised for the application of a more accurate and computationally expensive prediction model. We conclude that dynamic tracking of energy budgets combined with clustering could potentially be a powerful tool for turbulence analysis.

\FloatBarrier

\section{Conclusion and future work}
\label{sec:conclusions}

In this paper we have introduced a new clustering method BIA and its dynamic alternative. As previously emphasized, accurately determined clustering is crucial for model selection, i.e. a model with a complex PDE that accounts for all possible interactions, or a reduced model of the PDE at a given region.

The main contribution of the BIA algorithm is its ability to transform the feature space of each grid point so that it is characterised by equation-based attributes rather than input data. More specifically, each point is assessed based on the level of contribution of physical interaction terms from the underlying equation. A further contribution is the procedure for the dynamic evolution of the clusters. When applying the regression method to generate this transformation, we analysed the influence of the coordinate system to which the method should be applied. This conclusion is reasonable because the calculated equation should represent a particle and not a point in the spatial grid. For this purpose, a special input data pre-processing method compatible with the Python package PySINDy was developed. It tracks the particle motion and stores the time evolution of the variables for each particle in the Lagrangian coordinate system instead of the traditional Eulerian system. As the motion of each particle is stored during pre-processing, it was also possible to reconstruct the motion of the identified clusters. Thus, the dynamic evolution of the clusters was successfully captured. As mentioned above, the code developed is compatible with the open source Python PySINDy package, thus contributing to future research and development of the SINDy method.

This work has shown successful results and can be a promising starting point for the development of novel reliable software that could be coupled with commercial CFD solvers for dynamic identification and model selection purposes. Future research can consider the development of three-dimensional input pre-processing in the Lagrangian framework instead of the current two-dimensional alternative. This will allow stratified turbulence to be studied not only in the decay phase, where motion is mainly in the $x$-$y$ plane, but also by including interactions along the $z$-axis. In addition, parallelization of the solvers in the PySINDy package can be pursued. This can lead to speed improvements and the analysis of larger datasets. Finally, a dynamic strategy can be explored to combine the clustering proposed by the BIA algorithm with the model selection for each region in the numerical solver.

\section*{Acknowledgments}
We would like to acknowledge Dr. Nicolaos Petropoulos from University of Cambridge for helping with the dataset of stratified fluid flows, as well as for contributing to the guidance of this project. 

This research used resources of the Oak Ridge Leadership Computing Facility at the
Oak Ridge National Laboratory, which is supported by the Office of Science of the U.S. Department of Energy
under Contract No. DE-AC05-00OR22725.

\bibliography{references_paper.bib}

\newpage
\appendix

\section{The Turbulent Kinetic Energy (TKE) Equation}
\label{TKE_appendix_development}

The Turbulent Kinetic Energy (TKE) equation is defined as:
\begin{equation}
\begin{aligned}
    \overbrace{\pder{}{t} \left( \frac{1}{2} \ang{u_i' u_i'} \right)}^{\mathcal{K}'} &= 
    \overbrace{- 2 \nu \ang{s_{ij}' s_{ij}'}}^{\mathcal{E}'} - 
    \overbrace{\frac{g}{\rho_r} \ang{\rho' w'}}^{\mathcal{B}} +
    \overbrace{\left( -\ang{u_i' u_j'} S_{ij} \right)}^{\mathcal{P}} \\
    &\quad - \underbrace{\pder{}{x_i} \left[ \ang{p' u_i'} 
    + \ang{\left( u_i' + U_i \right) \frac{u_j' u_j'}{2}} 
    - \nu \left( \pder{}{x_i} \ang{\frac{u_j' u_j'}{2}} 
    + \pder{}{x_j} \ang{u_i' u_i'} \right) \right]}_{\nabla \cdot J}
\end{aligned}
\label{eq:tke_a}
\end{equation}

\noindent where \( p'\) is the pressure fluctuations, \( g \) gravity acceleration, \( \nu \) the kinematic viscosity, $s_{ij}' = \frac{1}{2}\left(\frac{\partial u_i'}{\partial x_j} + \frac{\partial u_j'}{\partial x_i}\right)$ and $S_{ij} = \frac{1}{2}\left(\frac{\partial U_i}{\partial x_j} + \frac{\partial U_j}{\partial x_i}\right)$ are respectively the perturbation strain-rate tensors and the mean strain-rate tensors.

\noindent Each term from the Eq.~\ref{eq:tke_a} should be expanded for \( i = 1, 2, 3 = x, y, z \) considering that \( u_x = u_1 = u \), \( u_y = u_2 = v \), and \( u_z = u_3 = w \). 
\\
\begin{itemize}
    \item The rate of irreversible energy lost, \( \mathcal{K}' \) is given by:
\end{itemize}

\begin{equation}
    \mathcal{K}' = \pder{}{t} \left( \frac{1}{2} \ang{u' u'} + \frac{1}{2} \ang{v' v'} + \frac{1}{2} \ang{w' w'} \right)
\end{equation}
\\
\begin{itemize}
    \item The dissipation rate of the turbulent kinetic energy, 
    \( \mathcal{E} \) is given by:
\end{itemize}

\begin{equation}
\begin{aligned}
    s_{xx}' &= \frac{1}{2} \left( \pder{u'}{x} + \pder{u'}{x} \right) = \pder{u'}{x} \quad  \quad \quad  \quad s_{xy}'=s_{yx}'= \frac{1}{2} \left( \pder{u'}{y} + \pder{v'}{x} \right),\\
    s_{yy}' &= \frac{1}{2} \left( \pder{v'}{y} + \pder{v'}{y} \right) = \pder{v'}{y} \quad \quad \quad  \quad s_{xz}'= s_{zx}'= \frac{1}{2} \left( \pder{u'}{z} + \pder{w'}{x} \right),\\
    s_{zz}' &= \frac{1}{2} \left( \pder{w'}{z} + \pder{w'}{z} \right) = \pder{w'}{z} \quad \quad \quad  \quad s_{yz}'=s_{zy}'= \frac{1}{2} \left( \pder{v'}{z} + \pder{w'}{y} \right)
\end{aligned}
\end{equation}

Substituting into the expression \( -2 \nu \langle s_{ij}' s_{ij}' \rangle \):
\begin{equation}
\begin{aligned}
    -2 \nu \left( 
    \underbrace{\ang{s_{xx}'^2}}_{\mathcal{E}_{xx}} + 
    \underbrace{\ang{s_{yy}'^2}}_{\mathcal{E}_{yy}} + 
    \underbrace{\ang{s_{zz}'^2}}_{\mathcal{E}_{zz}} + 
    \underbrace{2 \ang{s_{xy}'^2}}_{\mathcal{E}_{xy}} + 
    \underbrace{2 \ang{s_{xz}'^2}}_{\mathcal{E}_{xz}} + 
    \underbrace{2 \ang{s_{yz}'^2}}_{\mathcal{E}_{yz}} 
    \right)
\end{aligned}
\end{equation}

\begin{itemize}
    \item The rate of the turbulence production \( \mathcal{P} \) term is given by:
\end{itemize}

\begin{equation}
\begin{aligned}
    \mathcal{P}_{xx} &= -\ang{u' u'} \pder{U}{x}, \quad \quad \quad  \quad  \mathcal{P}_{xy} = \mathcal{P}_{yx} = -\ang{u' v'} \left( \frac{1}{2} \left( \pder{U}{y} + \pder{V}{x} \right) \right), \\
    \mathcal{P}_{yy} &= -\ang{v' v'} \pder{V}{y}, \quad \quad \quad  \quad \mathcal{P}_{xz} = \mathcal{P}_{zx} = -\ang{u' w'} \left( \frac{1}{2} \left( \pder{U}{z} + \pder{W}{x} \right) \right),\\
    \mathcal{P}_{zz} &= -\ang{w' w'} \pder{W}{z}, \quad \quad \quad  \quad  \mathcal{P}_{yz} = \mathcal{P}_{zy} = -\ang{v' w'} \left( \frac{1}{2} \left( \pder{V}{z} + \pder{W}{y} \right) \right)
\end{aligned}
\end{equation}

The previous terms are omitted in the calculations since \( U, V, W \) are the spatial averages that are considered constant for each variable over the subdomain.

\begin{itemize}
    \item Finally the divergence of the flux, \(J\) can be expanded as follows:
\end{itemize}

\begin{equation}
\begin{aligned}
    J_x &= - \pder{}{x} \Bigg[ \ang{p' u'} + \ang{\left( u' + U \right) \frac{u' u' + v' v' + w' w'}{2}} \\
    &\quad - \nu \left( \pder{}{x} \ang{\frac{u' u' + v' v' + w' w'}{2}} + \pder{}{x} \ang{u' u'} + \pder{}{y} \ang{u' u'} + \pder{}{z} \ang{u' u'} \right) \Bigg], \\
    J_y &= - \pder{}{y} \Bigg[ \ang{p' v'} + \ang{\left( v' + V \right) \frac{u' u' + v' v' + w' w'}{2}} \\
    &\quad - \nu \left( \pder{}{y} \ang{\frac{u' u' + v' v' + w' w'}{2}} + \pder{}{x} \ang{v' v'} + \pder{}{y} \ang{v' v'} + \pder{}{z} \ang{v' v'} \right) \Bigg], \\
    J_z &= - \pder{}{z} \Bigg[ \ang{p' w'} + \ang{\left( w' + W \right) \frac{u' u' + v' v' + w' w'}{2}} \\
     &\quad - \nu \left( \pder{}{z} \ang{\frac{u' u' + v' v' + w' w'}{2}} + \pder{}{x} \ang{w' w'} + \pder{}{y} \ang{w' w'} + \pder{}{z} \ang{w' w'} \right) \Bigg]
\end{aligned}
\end{equation}

From \(J\) the pressure term is omitted in the implementation since the flux non-divergence boundary condition is established. As well as \(J_x\), \(J_y\), \(J_z\) will be split into the following subterms:

\begin{equation}
\centering
\begin{aligned}
     J_{x_1} &= - \ang{\pder{}{x} \Bigg[ u' \frac{u'^2 + v'^2 + w'^2}{2} \Bigg]} \quad \quad \quad \quad
    J_{x_2} &= - \ang{\pder{}{x} \nu \left[ \pder{}{x} \left( \frac{u'^2}{2} \right) + \pder{}{x} (u'^2) \right]} \\
    J_{x_3} &= - \ang{\pder{}{x} \nu \left[ \pder{}{x} \left( \frac{v'^2}{2} \right) + \pder{}{y} (u'^2) \right]} \quad \quad
     J_{x_4} &= - \ang{\pder{}{x} \nu \left[ \pder{}{x} \left( \frac{w'^2}{2} \right) + \pder{}{z} (u'^2) \right]} \\
    J_{y_1} &= - \ang{\pder{}{y} \Bigg[ v' \frac{u'^2 + v'^2 + w'^2}{2} \Bigg]}, \quad \quad \quad \quad
    J_{y_2} &= - \ang{\pder{}{y} \nu \left[ \pder{}{x} \left( \frac{v'^2}{2} \right) + \pder{}{y} (v'^2) \right]} \\
    J_{y_3} &= - \ang{\pder{}{y} \nu \left[ \pder{}{x} \left( \frac{u'^2}{2} \right) + \pder{}{y} (v'^2) \right]}  \quad \quad
    J_{y_4} &= - \ang{\pder{}{y} \nu \left[ \pder{}{y} \left( \frac{w'^2}{2} \right) + \pder{}{z} (v'^2) \right]} \\
    J_{z_1} &= - \ang{\pder{}{z} \Bigg[ w' \frac{u'^2 + v'^2 + w'^2}{2} \Bigg]} \quad \quad \quad \quad
    J_{z_2} &= - \ang{\pder{}{z} \nu \left[ \pder{}{z} \left( \frac{w'^2}{2} \right) + \pder{}{y} (w'^2) \right]} \\
     J_{z_3} &= - \ang{\pder{}{z} \nu \left[ \pder{}{z} \left( \frac{u'^2}{2} \right) + \pder{}{x} (w'^2) \right]} \quad \quad
     J_{z_4} &= - \ang{\pder{}{z} \nu \left[ \pder{}{z} \left( \frac{v'^2}{2} \right) + \pder{}{y} (w'^2) \right]}
\end{aligned}
\end{equation}
\\

\section{The Turbulence Buoyancy Variance (TBV) Equation}
\label{TBV_appendix_development}

An alternative formulation of the TKE equation is the Turbulence Buoyancy Variance (TBV) equation written as :

\begin{equation}
\begin{aligned}
    \frac{1}{N^2} \pder{}{t} \ang{\frac{b^2}{2}} = 
    \underbrace{\frac{g}{\rho_r} \ang{\rho' w'}}_{\mathcal{B}} - 
    \underbrace{\frac{\kappa}{N^2} \ang{\pder{b}{x_i} \pder{b}{x_i}}}_{\chi} -
    \nabla \cdot \mathbf{J}_{\rho}   \\
    \nabla \cdot \mathbf{J}_{\rho} = 
    \sum_{i=1}^{3} \frac{1}{N^2} \pder{}{x_i} \left[ U_i \ang{\frac{b^2}{2}} + u'_i b' - \kappa b \pder{b}{x_i} \right]
\end{aligned}
\label{eq: tbv_a}
\end{equation}

where $b$ is buoyancy defined as $b = \rho' \frac{g}{\rho_r}$, $\kappa$ the molecular diffusivity and $N$ buoyancy frequency.\\

\noindent Each term from the Eq.~\ref{eq: tbv_a} should be expanded for \( i = 1, 2, 3 = x, y, z \) considering that \( u_x = u_1 = u \), \( u_y = u_2 = v \), and \( u_z = u_3 = w \). 
\\
\begin{itemize}
    \item The rate of buoyancy variance, \( \chi \) is given by:
\end{itemize}

\begin{equation}
\begin{aligned}
    \chi_x = \frac{\kappa}{N^2} \ang{\pder{b}{x} \pder{b}{x}} \quad \quad
    \chi_y = \frac{\kappa}{N^2} \ang{\pder{b}{y} \pder{b}{y}} \quad \quad
    \chi_z = \frac{\kappa}{N^2} \ang{\pder{b}{z} \pder{b}{z}}
\end{aligned}
\end{equation}

\begin{itemize}
    \item The transport flux, \( \mathbf{J}_{\rho} \) is given by:
\end{itemize}

\begin{equation}
\begin{aligned}
    J_x = \frac{1}{N^2} \pder{}{x} \left[ {   \underbrace{U \ang{\frac{b^2}{2}}}_{J_{x_0}} + \underbrace{u' b'}_{J_{x_1}} -  \underbrace{\kappa b \pder{b}{x}}_{J_{x_2}} } \right]  \quad \quad
    J_y = \frac{1}{N^2} \pder{}{y} \left[ \underbrace{V \ang{\frac{b^2}{2}}}_{J_{y_0}} + \underbrace{v' b'}_{J_{y_1}} - \underbrace{\kappa b \pder{b}{y}}_{J_{y_2}} \right] \\
    J_z =   \frac{1}{N^2} \pder{}{z} \left[ \underbrace{W \ang{\frac{b^2}{2}}}_{J_{z_0}} + \underbrace{w' b'}_{J_{z_1}} - \underbrace{\kappa b \pder{b}{z}}_{J_{z_2}} \right]
\end{aligned}
\end{equation}

\end{document}